\documentclass[%
 reprint,
superscriptaddress,
 amsmath,amssymb,
 aps,
]{revtex4-2}

\usepackage{graphicx}% Include figure files
\usepackage{dcolumn}% Align table columns on decimal point
\usepackage{bm}% bold math
\usepackage{comment}
\usepackage{physics}
\usepackage{footnote}
\usepackage{hyperref}

\begin{document}

\preprint{APS/123-QED}

\title{Optimization of Flight Routes: Quantum Approximate Optimization Algorithm for the Tail Assignment Problem}% Force line breaks with \\

\author{Marta Gili}
\email{mgili@ikerlan.es}
\affiliation{Ikerlan Technology Research Centre, Basque Research and Technology Alliance (BRTA), J.M. Arizmediarrieta Pasealekua 2, 20500, Arrasate-Mondragón, Spain}
\affiliation{Institute for Cross-Disciplinary Physics and Complex Systems (IFISC) UIB-CSIC, Campus Universitat Illes Balears, 07122,
Palma de Mallorca, Spain}%Lines break automatically or can be forced with \\

\author{Paul San Sebastian}%

\affiliation{Ikerlan Technology Research Centre, Basque Research and Technology Alliance (BRTA), J.M. Arizmediarrieta Pasealekua 2, 20500, Arrasate-Mondragón, Spain}%
\affiliation{
Universidad del País Vasco/Euskal Herriko Unibertsitatea UPV/EHU}%

\author{Ane Blázquez-García}
\affiliation{Ikerlan Technology Research Centre, Basque Research and Technology Alliance (BRTA), J.M. Arizmediarrieta Pasealekua 2, 20500, Arrasate-Mondragón, Spain}%

\begin{comment}
\collaboration{MUSO Collaboration}%\noaffiliation

\author{Charlie Author}
 \homepage{http://www.Second.institution.edu/~Charlie.Author}
\affiliation{
 Second institution and/or address\\
 This line break forced% with \\
}%
\affiliation{
 Third institution, the second for Charlie Author
}%
\author{Delta Author}
\affiliation{%
 Authors' institution and/or address\\
 This line break forced with \textbackslash\textbackslash
}%

\collaboration{CLEO Collaboration}%\noaffiliation
\end{comment}

\date{\today}% It is always \today, today,

\begin{abstract}

The Tail Assignment Problem (TAP) is a critical optimization challenge in airline operations, requiring the optimal assignment of aircraft to scheduled flights to maximize efficiency and minimize costs. To address the TAP, this work applies the Quantum Approximate Optimization Algorithm (QAOA), a promising quantum computing algorithm developed for tackling complex combinatorial optimization problems. A detailed formulation of the TAP is provided and QAOA's performance is evaluated on realistic problem instances, examining its strengths and weaknesses. Additionally, QAOA is compared with classical methods such as brute force and branch-and-price, as well as Quantum Annealing (QA), another quantum approach. The analysis reveals the current limitations of quantum hardware but suggests potential advantages as technology advances.

\begin{comment}
\begin{description}
\item[Usage]
Secondary publications and information retrieval purposes.
\item[Structure]
You may use the \texttt{description} environment to structure your abstract;
use the optional argument of the \verb+\item+ command to give the category of each item. 
\end{description}
\end{comment}

\end{abstract}

\maketitle

%\tableofcontents

\section{\label{sec:int}Introduction}

The Tail Assignment Problem (TAP) is a fundamental challenge in optimizing airline scheduling and logistics, aiming to efficiently assign aircraft (tails) to flights while minimizing costs and enhancing operational efficiency. This task involves managing complex constraints, such as flight connections and operational restrictions. In addition, the TAP plays an important role in crew scheduling and adherence to maintenance schedules. However, addressing this problem is particularly challenging due to its combinatorial complexity and NP-hard nature.
%Due to its combinatorial complexity and NP-hard nature, it is an ideal candidate for exploration.

Classical approaches to the TAP have been widely studied, among which is the branch-and-price algorithm. For instance, Grönkvist et al. explored a hybrid approach combining constraint programming with a branch-and-price algorithm \cite{gronkvist2005tail}. Ruther et al. also used a branch-and-price algorithm, but integrating the TAP with crew pairing \cite{ruther2017integrated}. Another applied classical approach has been Simulated Annealing (SA). Montoito et al. employed SA with an adaptive neighborhood local search to address this problem \cite{montoito2016application}.  

Although less commonly, quantum approaches have also been used to solve the TAP. These methods can be broadly divided into two groups, those using Quantum Annealing (QA) and those based on gate-based quantum algorithms. Martins et al. conducted the only known work applying QA to solve the TAP by developing a complex QUBO model \cite{martins2021qubo}. On the gate-based side, Vikstål et al. used the Quantum Approximate Optimization Algorithm (QAOA) to address the problem \cite{vikstaal2020applying}. 

QAOA is a hybrid quantum-classical algorithm designed for solving combinatorial optimization problems that has been recognized for its potential in various research studies. Farhi et al. initially demonstrated QAOA's applicability to combinatorial problems like Max-Cut and k-SAT, indicating its potential to outperform classical algorithms under specific conditions \cite{farhi2014quantum}. Further investigations by Zhou et al. underlined QAOA's robustness on Noisy Intermediate-Scale Quantum (NISQ) devices, where it achieved
near-optimal solutions despite hardware limitations \cite{zhou2020quantum}. Theoretical analyses by Farhi et al. explored QAOA's approximation guarantees, particularly affirming its potential computational advantage over classical methods \cite{farhi2016quantum}. Their work also postulated QAOA's role in demonstrating "quantum supremacy" under reasonable complexity-theoretic assumptions. Recent studies by Willsch et al. and Crooks et al. have examined QAOA's performance under realistic noise models and its implementation across diverse quantum hardware platforms, supporting the algorithm's promise and applicability in various contexts \cite{willsch2020benchmarking} \cite{crooks2018performance}.

Despite recent advancements, the application of QAOA to real-world problems, such as the TAP, remains limited. A key challenge includes handling the complexity of real-world instances, since they often involve large numbers of variables and constraints, making it difficult to efficiently explore the solution space \cite{omidvar2015designing}. In this context, another significant issue lies in the limitations of current quantum hardware, since NISQ devices pose difficulties in executing deep quantum circuits due to the significant error rates \cite{preskill2018quantum}. Additionally, refining strategies for optimizing QAOA parameters is essential, as finding the optimal parameters for each instance becomes increasingly difficult as the problem sizes grow \cite{zhou2020quantum}. Ongoing research focuses on addressing these challenges by optimizing QAOA's gate sequences, advancing in quantum error correction, and enhancing parameter tuning. Such improvements are crucial to overcome hardware limitations and unlock QAOA's full potential for large-scale optimization problems.

In the present work, the TAP is successfully solved using QAOA, taking into account route costs for a more accurate representation of the problem. This approach allows for new insights into the scalability and applicability of QAOA, extending the work previously introduced by Vikstål et al. \cite{vikstaal2020applying}. To further contextualize QAOA's performance, a comparative analysis is conducted using both classical methods, considering brute force and branch-and-price algorithms, as well as QA.

The rest of the paper is organized as follows. In Section 2, the proposed methodology is described. Section 3 provides the experimental results. Finally, Section 4 outlines the conclusions and future research lines.

\section{\label{sec:TAP}The Tail Assignment Problem}

The TAP is a central challenge in the airline industry, where the task is to assign specific aircraft (tails) to a series of scheduled flights. The primary objective is to minimize operational costs while satisfying a range of complex constraints. These include ensuring proper flight connections, adhering to maintenance schedules, and meeting flight restrictions, as not every aircraft is suitable for all routes. Effectively solving this problem can result in substantial cost reductions, improved operational reliability, and more efficient resource utilization.

This study examines the TAP with a focus on minimizing operational costs while satisfying flight connection constraints. To address this problem, it is essential to define the concept of route, which refers to a sequence of flights operated consecutively by the same aircraft. For a route to be considered operationally viable, one flight must arrive at the destination airport before another flight departing from the same airport, ensuring a minimal turnaround time between these consecutive flights, which can be set to $1$ hour. 

Based on Vikstål et al., the TAP can be expressed as:

\begin{comment}
\begin{equation}
    min \hspace{0.18cm} \sum_{r \in R} c_r x_r + \sum_{f \in F} C_f u_f     
\end{equation}

\begin{equation}
    subject  \hspace{0.14cm} to  \hspace{0.18cm} \sum_{r \in R} a_{fr} x_r + u_f = 1 \hspace{0.25cm} \forall f \in F
    \label{eq:c1}
\end{equation}

\begin{equation}
    \hspace{1.58cm} \sum_{r \in R} b_{tr} x_r + v_t = 1 \hspace{0.25cm} \forall t \in T
    \label{eq:c2}
\end{equation}
\end{comment}

\begin{eqnarray}
     min \hspace{0.18cm} \sum_{r \in R} c_r x_r + \sum_{f \in F} C_f u_f \label{eq:c} \\
     subject  \hspace{0.1cm} to  \hspace{0.18cm} \sum_{r \in R} a_{fr} x_r + u_f = 1 \hspace{0.25cm} \forall f \in F \label{eq:c1} \nonumber \\
    \hspace{1.58cm} \sum_{r \in R} b_{tr} x_r + v_t = 1 \hspace{0.25cm} \forall t \in T \label{eq:c2} \nonumber  \\
    x_r, u_f, v_t \in \{0,1\} \nonumber \\ \nonumber
\end{eqnarray}

where $F$ is the set of flights, $T$ the set of tails, and $R$ the set of valid routes \cite{vikstaal2020applying}. The vector of constants $c_r$ represents the cost of operating each route $r$, while $C_f$ stands for the cost of leaving flight $f$ unassigned. The matrix $a_{fr}$ represents which flights are covered by each route, so that $a_{fr}=1$ if flight $f$ is covered by route $r$, and $0$ otherwise. On the other hand, $b_{tr}$ relates routes to tails, such that $b_{tr}=1$ if route $r$ is performed by tail $t$, and $0$ otherwise. For the decision variables, $x_r=1$ if route $r$ is used in the solution, $u_f=1$ if flight $f$ is left unassigned, and $v_t=1$ if tail $t$ is unused, and $0$ otherwise.

The objective is to minimize the total cost, adhering to various constraints. The first constraint ensures that each flight is present in exactly one route of the solution, while the second constraint ensures that each tail is used at most once. For simplicity, other constraints, such as maintenance constraints, are not considered in this approach.

\section{\label{sec:met}Methodology}

In this section, first, a detailed explanation of the used optimization algorithm (QAOA) is presented (see Section \ref{sec:QAOA}), followed by the formulation proposed in this paper to address the TAP (see Section \ref{sec:form}).

\subsection{\label{sec:QAOA}The Quantum Approximate Optimization Algorithm}

QAOA belongs to the class of Variational Quantum Algorithms (VQAs) and it is inspired by Adiabatic Quantum Computation (AQC). Unlike in AQC, where the system is evolved through a slow adiabatic process, QAOA discretizes the evolution into parameterized quantum circuits. 

\begin{figure}[h]
    \centering
    \includegraphics[scale=0.17]{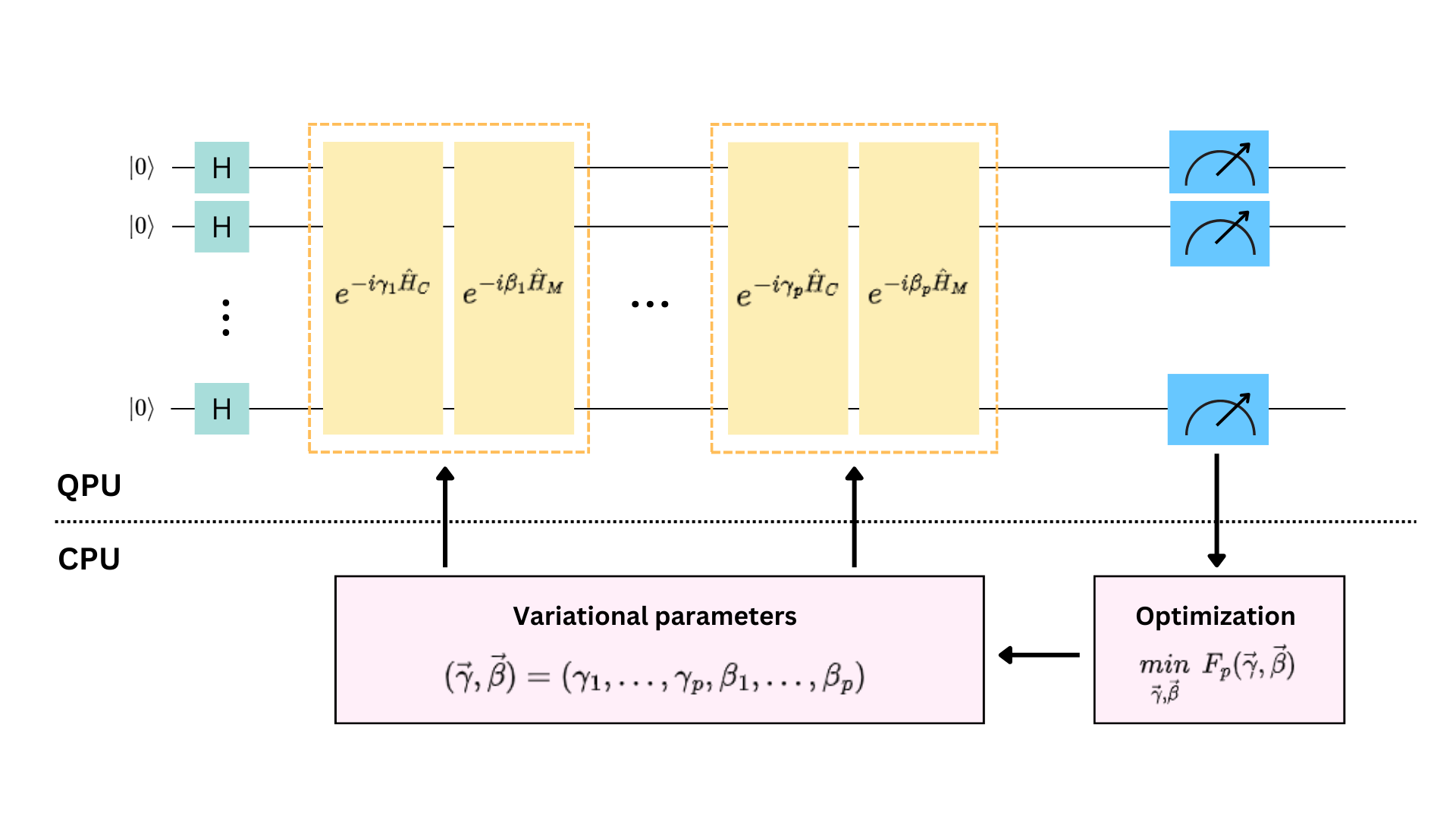}
    \caption[Diagram of the workflow of QAOA with p layers]{Diagram of the workflow of QAOA with $p$ layers.}
    \label{fig:circ}
\end{figure}

A diagram of the workflow of QAOA is shown in Figure \ref{fig:circ}. It illustrates a hybrid quantum-classical process, where the quantum part begins by initializing qubits and applying a sequence of quantum gates with adjustable parameters to evolve the quantum state. Then, after measurement, the classical part updates these parameters using optimization methods. This iterative cycle between the quantum and classical components continues until the optimal solution is found. 

An important part of the implementation of the algorithm is to translate the objective function $C(x)$ of the optimization problem into its quantum analog, the cost Hamiltonian ($\hat{H}_C$), which is defined such that its ground state encodes the corresponding solution (Eq. \ref{eq:Hc}). Here, $x \in \{0,1\}^n$ represents a binary bitstring, where each element corresponds to a variable in the optimization problem.

\begin{equation}
    \hat{H}_C \ket{x} = C(x) \ket{x}
    \label{eq:Hc}
\end{equation}

Additionally, a mixer Hamiltonian ($\hat{H}_M$) is defined to explore the solution space. It is typically chosen as $\hat{H}_M=\sum_{i=1}^n \hat{\sigma}_i^x$, although alternative formulations have also been explored \cite{hadfield2019quantum}. The circuit initialization involves preparing a superposition of the eigenstates of $\hat{H}_M$, achieved in this case by applying Hadamard gates to the initial $\ket{0}^{\otimes n}$ state, to create $\ket{+}^{\otimes n}$. 

The circuit ansatz is constructed using the unitary evolution operators $\hat{U}_C(\gamma)$ and $\hat{U}_M(\beta)$, defined in Eq. \ref{eq:uc_um}. Here, $\gamma \in [0,2\pi]$ (if $\hat{H}_C$ has integer-valued eigenvalues) and $\beta \in [0,\pi]$ represent the variational parameters of the circuit.

%$\gamma \in [0, 2\pi]$ (if the eigenvalues of $\hat{H}_C$ are integer-valued) and $\beta \in [0, \pi]$. 

\begin{equation}
    \hat{U}_C(\gamma) = e^{-i\gamma \hat{H}_C}, \quad \hat{U}_M(\beta) = e^{-i\beta \hat{H}_M}
    \label{eq:uc_um}
\end{equation}

These unitaries are known as cost and mixer layers and they can be applied consecutively $p$ times, resulting in $2p$ parameters: $\vec{\gamma}=(\gamma_1,\gamma_2,...,\gamma_p)$ and $\vec{\beta}=(\beta_1,\beta_2,...,\beta_p)$. The state at the output of the circuit is given by Eq. \ref{eq:ev}.
\begin{equation}
\ket{\psi(\vec{\gamma},\vec{\beta})}= \hat{U}_M(\beta_p)\hat{U}_C(\gamma_p)...\hat{U}_M(\beta_1)\hat{U}_C(\gamma_1) \ket{+}^{\otimes n}
    \label{eq:ev}
\end{equation}

Parameters are initialized randomly or using other techniques (e.g. heuristic strategies for finding quasi-optimal parameters efficiently, layer-by-layer optimization, warm-start techniques using GNNs, parameter transferability across graphs, or initialization using Trotterized Quantum Annealing) and then iteratively updated through classical gradient descent optimization, aiming to minimize the expectation value of $\hat{H}_C$ with respect to the output state of the circuit (Eq. \ref{eq:exp} and Eq. \ref{eq:min}). Other classical optimizers, such as Nelder-Mead, COBYLA, or BFGS, can also be used, depending on the specific characteristics of the problem and the optimization landscape \cite{blekos2024review}.

\begin{equation}
F_p(\vec{\gamma},\vec{\beta})=\bra{\psi(\vec{\gamma},\vec{\beta})} \hat{H}_C \ket{\psi(\vec{\gamma},\vec{\beta})}
    \label{eq:exp}
\end{equation}

\begin{equation}
    (\vec{\gamma}^*,\vec{\beta}^*) = arg \hspace{0.1cm} \underset{\vec{\gamma},\vec{\beta}}{min} \; F_p(\vec{\gamma},\vec{\beta})
    \label{eq:min}
\end{equation}

Once the parameters are optimized, sampling at the circuit output provides probabilities for each bitstring, which encode the problem's solution. The bitstring with the highest probability will be selected as the potential optimal solution.

\subsubsection{Trotterization}

The total Hamiltonian is expressed as the sum of the cost and mixer Hamiltonians ($\hat{H}=\hat{H}_M+\hat{H}_C$), which do not necessarily commute. Then, to apply a full layer in the circuit, the Trotter-Suzuki decomposition formula in Eq. \ref{eq:trot} is required, from which the corresponding unitary operator is approximated by Eq. \ref{eq:app}.

\begin{equation}
    e^{A+B}=\lim_{n \to \infty} \left( e^{A/n} e^{B/n} \right)^n
    \label{eq:trot}
\end{equation}

\begin{equation}
    \hat{U}(t) = e^{-it\hat{H}(t)} \approx \left( e^{-it\hat{H}_M/n} e^{-it\hat{H}_C/n} \right)^n
    \label{eq:app}
\end{equation} 

Note that the larger $n$, which is known as Trotterization depth, the better the approximation \citep{suzuki1976generalized}. Therefore, applying more layers is expected to improve the approximation of the solution, increasing the probability of finding the optimal solution (success probability) as the number of layers grows.

In fact, QAOA can approximate Adiabatic Quantum Annealing (AQA) through Trotterization for large $p$. In AQA, the evolution from the initial to the final Hamiltonian must be sufficiently slow to ensure the system remains in its ground state, thus leading to the correct solution. In QAOA, the circuit is initialized in the ground state of the initial Hamiltonian ($\hat{H}_M$) and evolves towards the ground state of the final Hamiltonian ($\hat{H}_C$), where the parameters $\gamma$ and $\beta$ represent the time coefficients that guide this evolution. Consequently, repeatedly applying layers in QAOA is equivalent to repeatedly applying small time steps (Trotter steps) to approximate the adiabatic evolution. 

\subsubsection{Parameter initialization}

As discussed in Zhou et al., the task of finding quasi-optimal parameters through random initialization requires $2^{O(p)}$ optimization runs, which becomes exponentially difficult as $p$ increases \cite{zhou2020quantum}. Addressing this issue, they introduced heuristic optimization strategies for large $p$, based on observed patterns in the optimal parameters using the BFGS algorithm applied to the Max-Cut problem across different graph instances. These observations facilitate the selection of educated guesses of variational parameters at the $(p+1)$-level based on optimized parameters at level $p$. The proposed strategies do not guarantee finding the global optimum, but they have been shown to efficiently find quasi-optimal solutions in $O(poly(p))$ time.

One of them is the Interpolation-based strategy (INTERP). This method uses linear interpolation to generate a good initial point as $p$ increases. The idea is to begin with $p=1$ and increment $p$ after obtaining a local optimum. The parameter relations are shown in Eq. \ref{eq:g} and Eq. \ref{eq:b}, where $i=1,2,...,p$. In these expressions, $(\gamma)_i$ and $(\beta)_i$ are the \textit{i}-th element of $\vec{\gamma}$ and $\vec{\beta}$, and $(\gamma_{p}^L)_{0}=(\gamma_{p}^L)_{p+1}=(\beta_{p}^L)_{0}=(\beta_{p}^L)_{p+1}=0$.

\begin{equation}
    (\gamma_{(p+1)}^0)_i = \frac{i-1}{p} (\gamma_{p}^L)_{i-1} + \frac{p-i+1}{p} (\gamma_{p}^L)_{i}
    \label{eq:g}
\end{equation}

\begin{equation}
    (\beta_{(p+1)}^0)_i = \frac{i-1}{p} (\beta_{p}^L)_{i-1} + \frac{p-i+1}{p} (\beta_{p}^L)_{i}
    \label{eq:b}
\end{equation}

Starting from a local optimum $\left( \gamma_{p}^L, \beta_{p}^L \right)$, a suitable initial point $\left( \gamma_{p+1}^0, \beta_{p+1}^0 \right)$ is obtained, from which the next local optima $\left( \gamma_{p+1}^L, \beta_{p+1}^L \right)$ can be found, and the process continues iteratively.

At the beginning, combinations of $\gamma$ and $\beta$ for $p=1$ are evaluated to determine the minimum expectation value across a grid of possible configurations. It is essential to account for any symmetry in the parameter space to simplify the process. If the eigenvalues of both the cost and mixer Hamiltonians are integer-valued, the expectation value exhibits even symmetry: $F_p(\vec{\gamma},\vec{\beta})=F_p(-\vec{\gamma},-\vec{\beta})$. This symmetry allows $\gamma$ to be restricted to the range $\gamma \in [0,\pi]$. 

\begin{figure}[h!]
    \centering
    \includegraphics[scale=0.25]{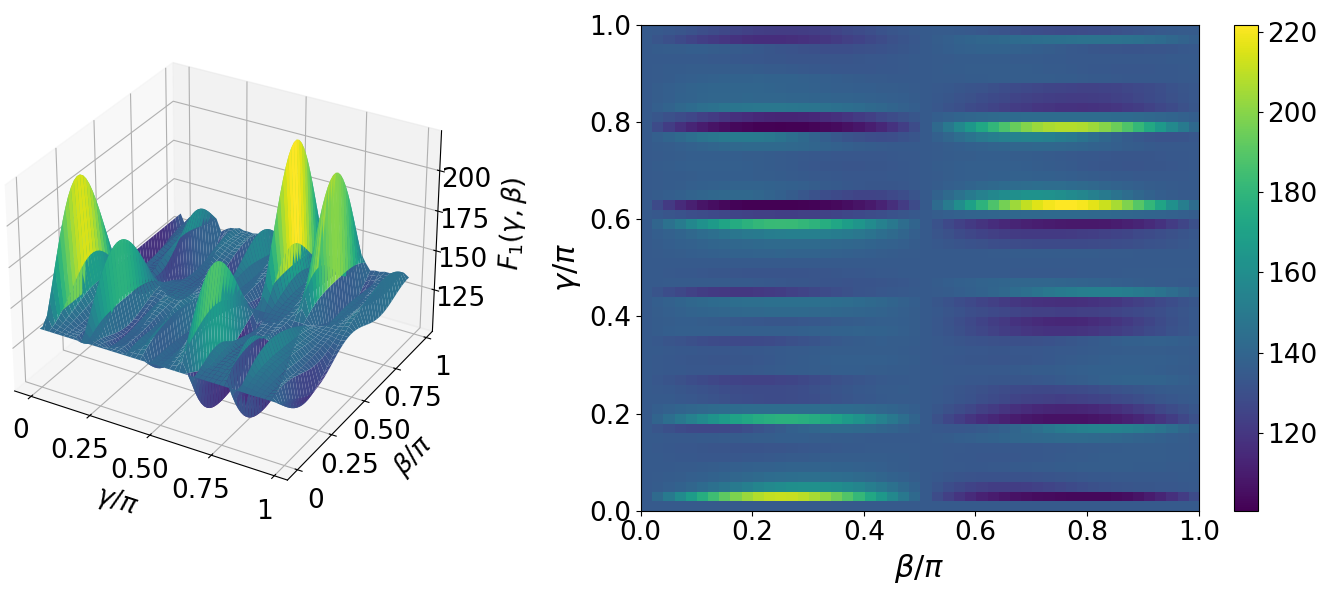}
    \caption{Example of the expectation value of the cost Hamiltonian as a function of the variational parameters $\gamma$ and $\beta$ for $p=1$.}
    \label{fig:en_land}
\end{figure}

Figure \ref{fig:en_land} shows an example of the expectation value of the cost Hamiltonian at the output of the circuit as a function of the variational parameters $\gamma$ and $\beta$ for $p=1$. From this plot, the parameters that yield the minimum expectation value can be identified and used as the initial $\left( \gamma_{1}^L, \beta_{1}^L \right)$. Subsequently, the INTERP heuristic is applied to find a suitable starting point for $p=2$, obtain a local optimum, and repeat this process iteratively for increasing values of $p$. 

It is noteworthy that as the landscape becomes more complex, more challenges will arise in finding the optimal solution to the problem.

\subsection{\label{sec:form}Formulation of the Problem}

In our scenario, it is assumed that there exists a solution where all flights are assigned and therefore the variable $u_f$ can be removed from the model. Furthermore, it is supposed that there is a sufficient number of aircraft available to assign one to each route within the solution. These aircraft are considered suitable for any route, implying full availability across all routes in the solution. Then, the second constraint defined in Eq. \ref{eq:c} can be removed and the model is reduced to: 

\begin{comment}
\begin{equation}
    min \hspace{0.18cm} \sum_{r \in R} c_r x_r 
    \label{eq:min2}
\end{equation}

\begin{equation}
    subject  \hspace{0.1cm} to  \hspace{0.18cm} \sum_{r \in R} a_{fr} x_r = 1 \hspace{0.25cm} \forall f \in F
    \label{eq:c12}
\end{equation}
\end{comment}

\begin{eqnarray}
     min \hspace{0.18cm} \sum_{r \in R} c_r x_r  \\
     subject  \hspace{0.1cm} to  \hspace{0.18cm} \sum_{r \in R} a_{fr} x_r = 1 \hspace{0.25cm} \forall f \in F \nonumber \\
    x_r \in \{0,1\} \nonumber \\ \nonumber
\end{eqnarray}

As a novelty compared to Vikstål et al. the costs associated with the routes are taken into account, rather than considering only the decision version of the problem in which the goal is to find any solution satisfying the constraint (with $min \hspace{0.18cm} \sum_{r \in R} c_r x_r$ reduced to $min \hspace{0.18cm} 0$) \cite{vikstaal2020applying}.

In terms of a Quadratic Unconstrained Binary Optimization (QUBO) problem, the objective function to be minimized is given by Eq. \ref{eq:qubo}.

\begin{equation}
    Q(x)=  \sum_{r \in R} c_r x_r + P \sum_{f \in F} \left( 1 - \sum_{r \in R} a_{fr} x_r \right)^2
    \label{eq:qubo}
\end{equation}

The constraint is incorporated as a penalty term with a weight determined by $P$. This term imposes higher energy contributions for any states that violate the constraint, thereby discouraging such configurations.

\subsubsection{Ising Hamiltonian}

Applying QAOA involves encoding the objective function into a Hamiltonian, which represents the total energy of the quantum system. An effective qubit encoding scheme is essential to translate the binary decision variables to quantum computation. In this context, the Ising model (Eq. \ref{eq:ising}) is used, where Pauli-Z operators are introduced. 

\begin{equation}
    x_{r} \rightarrow \frac{(1-\hat{\sigma}_r^z)}{2}
    \label{eq:ising}
\end{equation}

This leads to the cost Hamiltonian in the form of Eq. \ref{eq:hc}, from where it can be seen that to solve this problem a quantum processor of $r$ qubits is needed.

\begin{equation}
    \hat{H}_C = \sum_r h_r \hat{\sigma}_r^z + \sum_{r<r'} J_{rr'} \hat{\sigma}_r^z \hat{\sigma}_{r'}^z + const
    \label{eq:hc}
\end{equation}

After calculation, the coefficients are given by:

\begin{equation}
    h_r = -\frac{1}{2} c_r - \frac{P}{2} \sum_f a_{fr} \left( \sum_{r'} a_{fr'} - 2 \right)
\end{equation}

\begin{equation}
    J_{rr'} = \frac{P}{4}  \sum_f a_{fr} a_{fr'}
\end{equation}

\begin{equation}
   const = \frac{1}{2} \sum_r c_r + \frac{P}{4} \sum_f  \left( \sum_{r} a_{fr} - 2 \right)^2
\end{equation}

When constructing the circuit with the unitaries based on the cost and mixer Hamiltonians, the operations translate into X-rotation gates for $\hat{U}_M$, and Z-rotation gates and Ising ZZ-coupling gates for $\hat{U}_C$. The Ising ZZ-coupling gates are implemented using a combination of two CNOT gates and a Z-rotation gate.

\section{\label{sec:exp}Experimentation}

This section presents the experimentation process, detailing the used dataset (see Section \ref{sec:data}) and the results obtained from applying QAOA to the generated instances (see Section \ref{sec:res}), along with a comparison of its performance against other approaches (see Section \ref{sec:comp}).

\subsection{\label{sec:data}Dataset and Instances}

For this work, the selected open dataset was chosen due to its accessibility and frequent updates, in addition to providing the data necessary to generate the instances \footnote{\url{https://www.transtats.bts.gov/DL_SelectFields.aspx?gnoyr_VQ=FGJ&QO_fu146_anzr=b0-gvzr}}.

This dataset is sourced from the Bureau of Transportation Statistics (BTS), an arm of the United States Department of Transportation (DOT). Specifically, is part of the Airline On-Time Performance Data. It contains the date of the flights, origin and destination airports, and departure and arrival times reported by certified U.S. air carriers, among other characteristics that have not been taken into account for this work.
%\footnote{https://www.transtats.bts.gov/DatabaseInfo.asp?QO\_VQ \\ =EFD\&Yv0x=D}

The date January 1, 2024 was taken, for which the dataset includes 17,265 flights. Then, several instances of different sizes and complexity were considered to comprehensively evaluate the performance of QAOA across a range of real scenarios. 

The first step was to select a subset of flights and search for all possible connections between them to form valid routes. However, the instances were limited to at most 15 routes, corresponding to 15 qubits, as this is close to the maximum computational capacity of quantum device simulations. As computational resources continue to advance, the same principles and methods can be applied to solve larger and more complex instances. 

The cost of each route in a problem instance was determined based on values provided by the International Civil Aviation Organization (ICAO)\footnote{\url{https://www.icao.int/Pages/default.aspx}}. DOT Form 41 provides financial information about U.S. major airlines, specifically their airline operating costs. These costs include crew, fuel, maintenance, and ownership expenses, allowing to calculate the total cost per block hour. Block time refers to the time of a flight measured from the start of movement out of the parking position to the end of movement, which includes both flight time and taxi time at the airport. Table \ref{tab:costs} contains this information.

\begin{table}[h!]
\caption{\label{tab:costs}%
Airline operating costs per block hour.
}
\begin{ruledtabular}
\begin{tabular}{lcdr}
\textrm{\textbf{Cost Type}}&
\textrm{\textbf{Amount (\$ per block hour)}}\\
\colrule
Crew Costs  & 489\\
Fuel Costs  & 548\\
Maintenance Costs  & 590\\
Ownership Costs  & 923\\
\hline
\textbf{Total}  & \textbf{2,550}\\
\end{tabular}
\end{ruledtabular}
\end{table}

By knowing the block time of each flight, the total cost of a route could be calculated. Additionally, an extra cost for routes with fewer flights was added, as this results in aircraft spending more time on the ground, which is not optimal.ection{\label{sec:res}Results applying QAOA}

The results of applying QAOA to the created real-world instances are outlined below, with the analysis divided into two main parts. The first part evaluates the algorithm's ability to distinguish the optimal solution from the set of all feasible solutions. The second part explores the effect of the route graph connectivity on the problem complexity. 

In the route graph, nodes represent routes, and edges connect routes that share common flights. According to Vikstål et al., the number of edges incident to a node, known as valency, plays a crucial role in the problem's difficulty \cite{vikstaal2020applying}. A larger average valency increases the problem's complexity since each flight can be selected only once. Consequently, the more routes a flight appears in, the more challenging it becomes to determine which routes to choose. For this reason, instances with an average valency below 2 were considered.

\subsubsection{Optimal vs Suboptimal Solutions}

\begin{figure}[h!]
    \centering
    \includegraphics[width=0.7\linewidth]{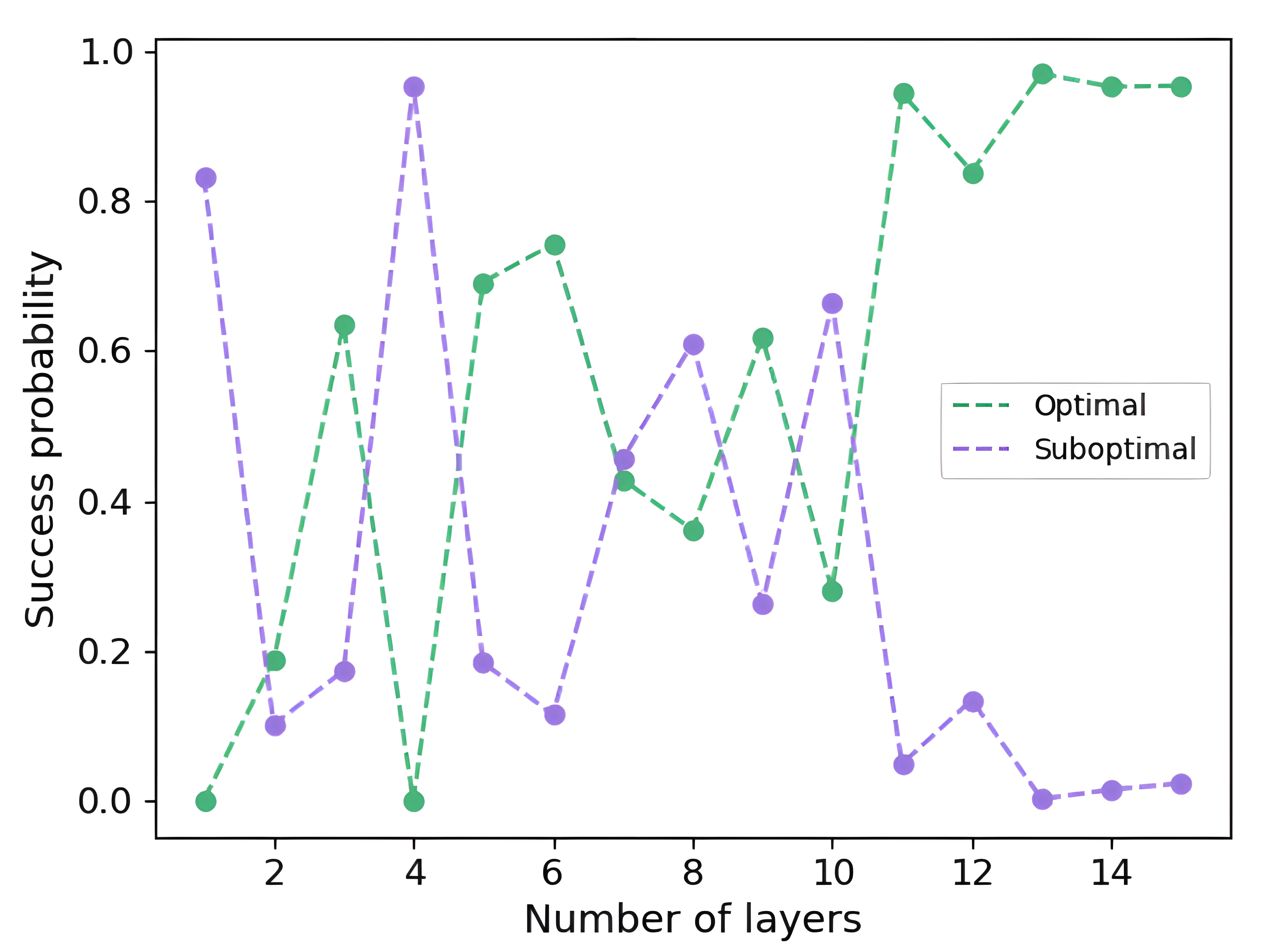}
    \caption[Optimal and suboptimal solution for an instance of 6 routes]{Balance between the optimal and the suboptimal solution for an instance of 6 routes and average valency of 0.66.}
    \label{fig:subopt}
\end{figure}

The algorithm can generally find feasible solutions, but identifying the optimal among them is more challenging. This is illustrated in Figure \ref{fig:subopt}, where the INTERP strategy was applied to an instance of 6 routes and a low average valency of 0.66 in the route graph. In this example, there are only two feasible solutions: one optimal and one suboptimal. A suboptimal solution, in this context, is one that covers all routes exactly once but does not have the lowest cost. 
    
Initially, the algorithm oscillates between the optimal and suboptimal solutions. However, from 11 layers onward, it converges towards the optimal solution.

\begin{figure}[h!]
    \centering
    \includegraphics[width=0.70\linewidth]{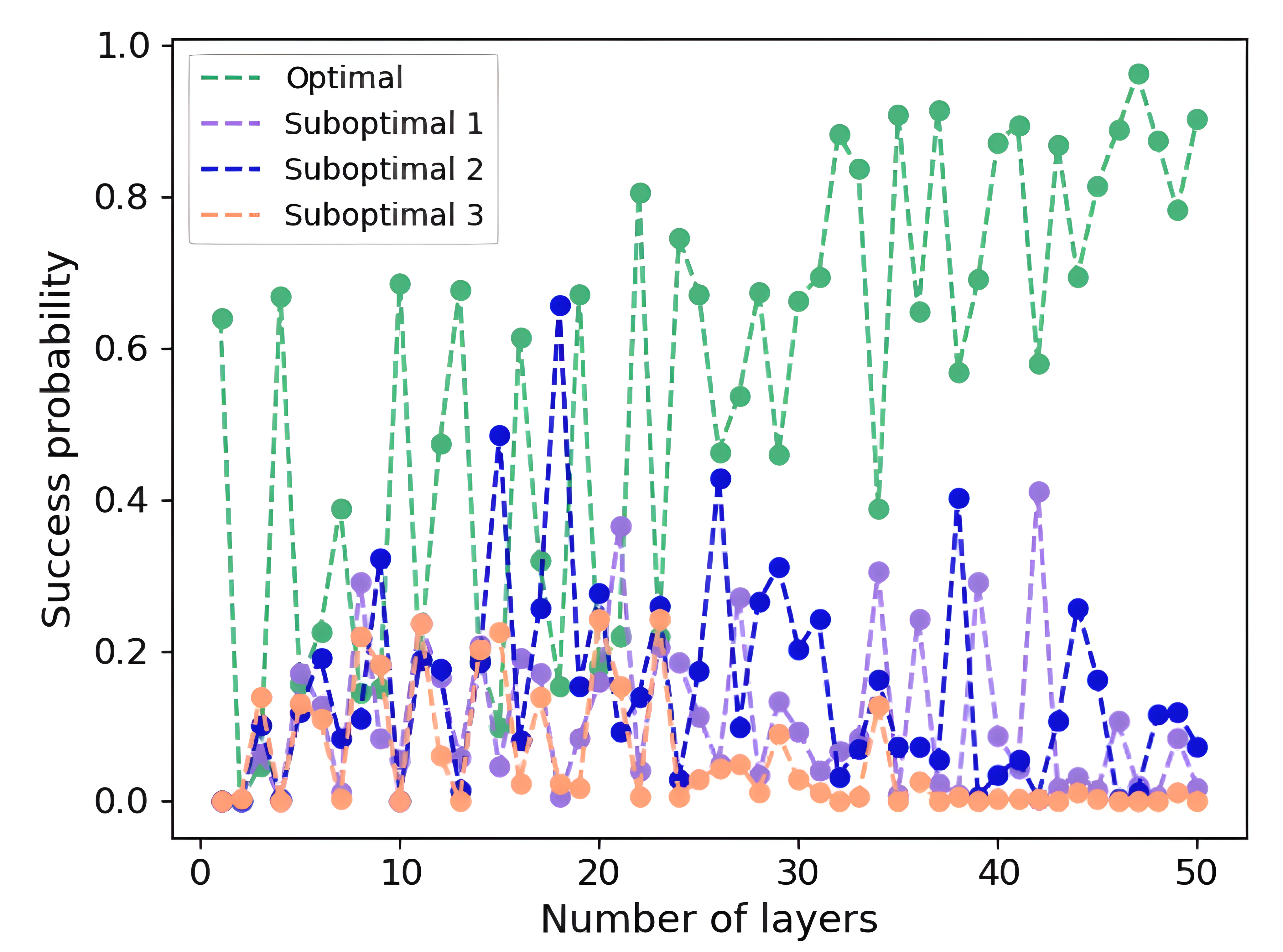}
    \caption[Optimal and suboptimal solution for an instance of 7 routes]{Balance between the optimal and the suboptimal solutions for an instance of 7 routes and average valency of 1.14.}
    \label{fig:subopt2}
\end{figure}

The same happens if there is more than one suboptimal solution, as in the case of Figure \ref{fig:subopt2}, for an instance of 7 routes and an average valency of 1.14. Now, the algorithm needs more layers to finally opt for the best solution, due to the increase of complexity of the problem.

\subsubsection{Connectivity in the route graph}

The effect of the route graph connectivity was analyzed for instances of 6 and 10 routes, with different average valencies. The INTERP strategy was used for the results, executing each instance 5 times to achieve more reliable results.

\begin{figure}[h!]
    \centering
    \begin{minipage}{0.145\textwidth}
        \centering
        \includegraphics[width=\textwidth]{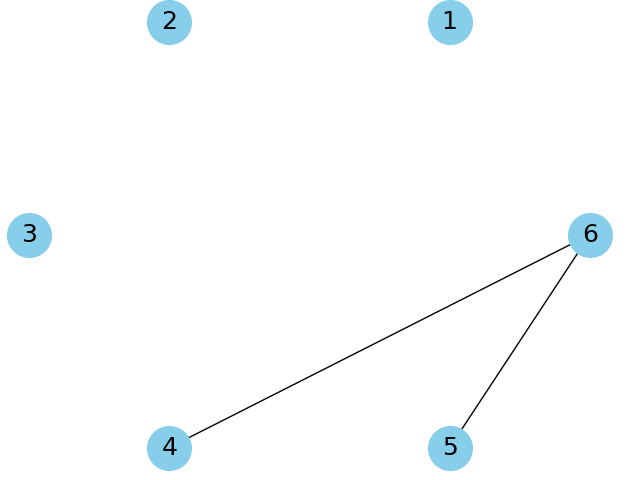}
    \end{minipage}\hfill
    \begin{minipage}{0.15\textwidth}
        \centering
        \includegraphics[width=\textwidth]{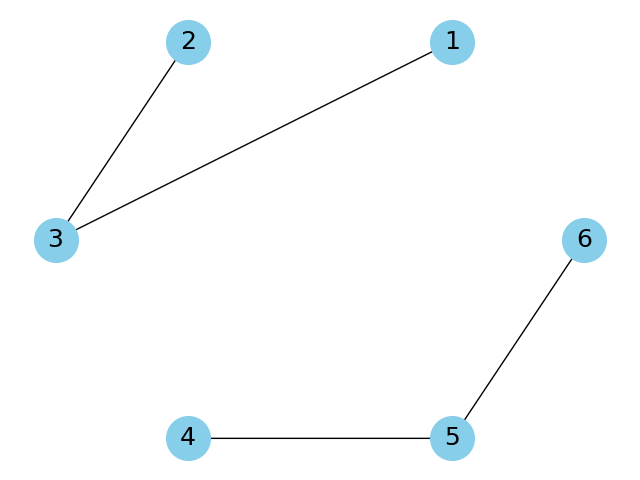}
    \end{minipage}\hfill
    \begin{minipage}{0.15\textwidth}
        \centering
        \includegraphics[width=\textwidth]{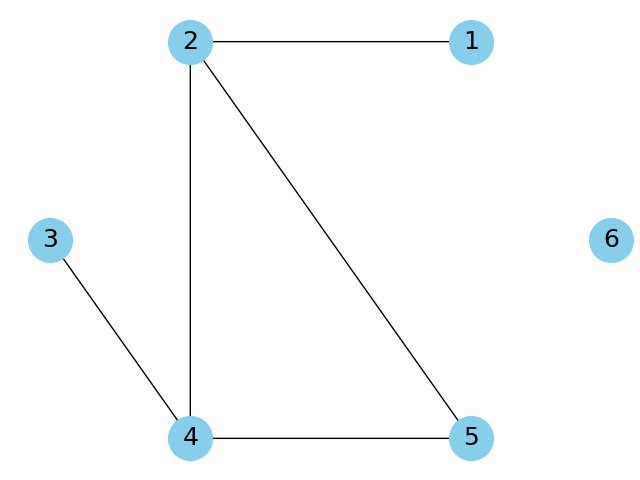}
    \end{minipage}
    \centering
    \includegraphics[width=0.75\linewidth]{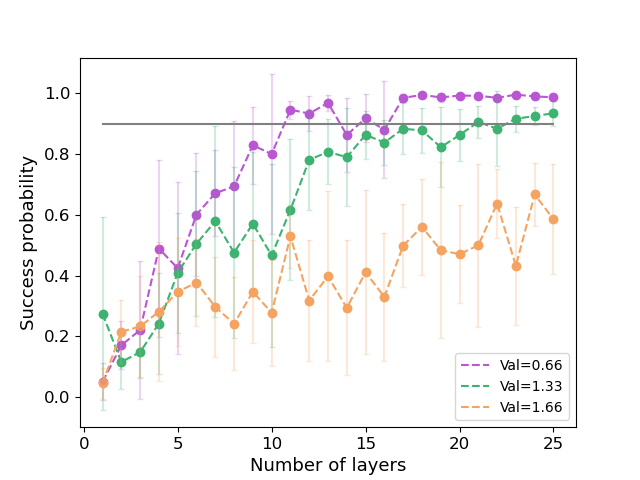}
    \caption{Top: Route graphs with average valencies of 0.66 (left), 1.33 (middle) and 1.66 (right). Bottom: Success probability when considering different average valencies for an instance of 6 routes.}
    \label{fig:gr1}
\end{figure}

For the instance of 6 routes, average valencies of 0.66, 1.33, and 1.66 were considered. Each route graph and the corresponding success probability adding up to 25 layers are shown in Figure \ref{fig:gr1}. It can be seen that as the average valency increases, getting the optimal solution with high probability becomes more difficult.

\begin{figure}[h!]
    \centering
    \begin{minipage}{0.15\textwidth}
        \centering
        \includegraphics[width=\textwidth]{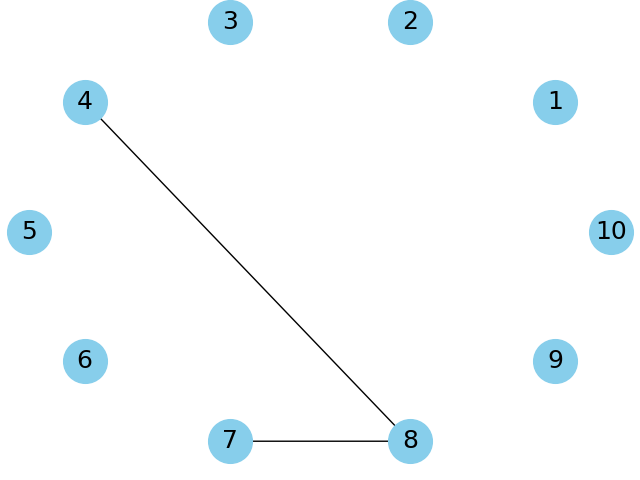}
    \end{minipage}\hfill
    \begin{minipage}{0.15\textwidth}
        \centering
        \includegraphics[width=\textwidth]{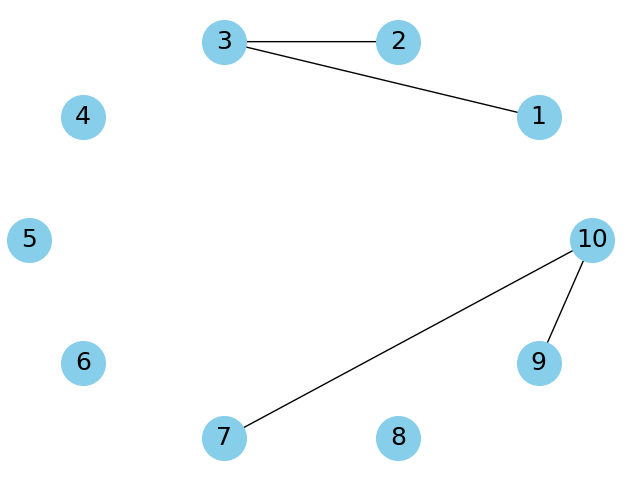}
    \end{minipage}\hfill
    \begin{minipage}{0.15\textwidth}
        \centering
        \includegraphics[width=\textwidth]{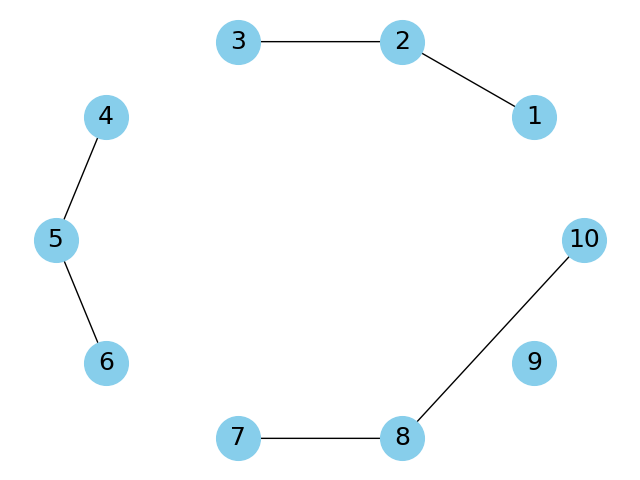}
    \end{minipage}
    \includegraphics[width=0.75\linewidth]{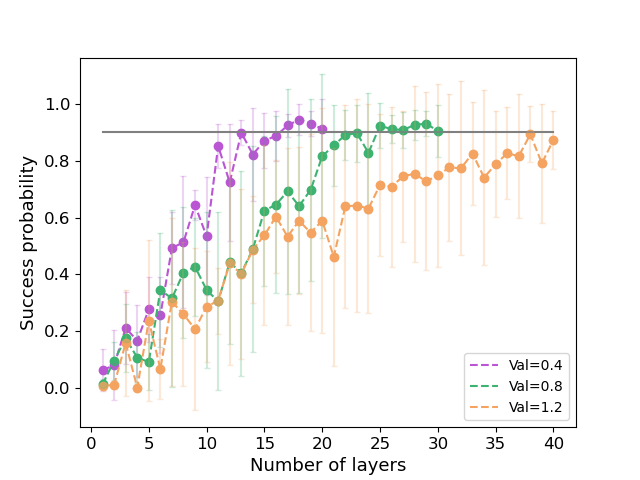}
    \caption{Top: Route graphs with average valencies of 0.4 (left), 0.8 (middle) and 1.2 (right). Bottom: Success probability when considering different average valencies for an instance of 10 routes.}
    \label{fig:gr2}
\end{figure}

Instead of fixing a maximum number of layers, layers can be added until a success probability close to 90\% is found. This is what was done for the instances of 10 routes, with average valencies of 0.4, 0.8, and 1.2 (Figure \ref{fig:gr2}). As before, increasing the average valency makes it more challenging to find the solution.

\vspace{1cm}

\begin{figure}[h!]
    \centering
    \includegraphics[width=0.9\linewidth]{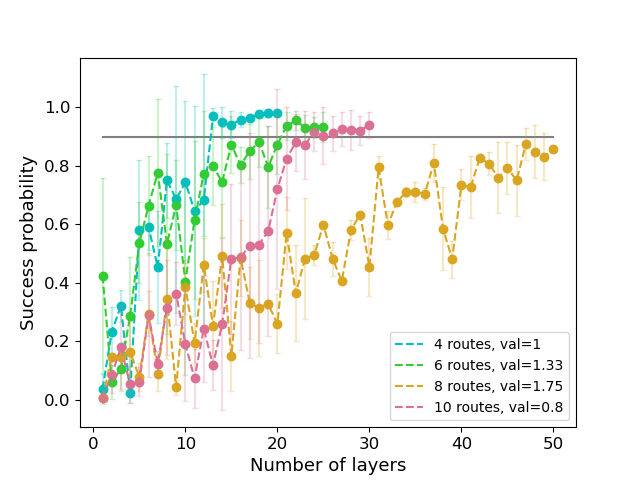}
    \caption{Success probability when considering different numbers of routes and different average valencies.}
    \label{fig:comp}
\end{figure}

As a final consideration, instances with different number of routes, apart from different average valencies, were examined. In general, increasing the number of routes, equivalent to adding more qubits, results in a more complex problem. This is further accentuated by higher connectivity within the route graph. In Figure \ref{fig:comp}, where a sufficient number of layers to achieve a success probability of the 90\% were applied, the instances with 4, 6, and 8 routes, and increasing average valencies of 1, 1.33, and 1.75, illustrate this trend. However, an instance with 10 routes but a lower average valency of 0.8 becomes easier to solve than the one with 8 routes. This indicates that the problem complexity is influenced not only by the number of routes (qubits) but also by the connectivity within the route graph.

\subsection{\label{sec:comp}Comparative Analysis with Alternative Approaches}

Benchmarking quantum and classical algorithms presents significant challenges, as discussed in Bucher et al. \cite{bucher2024robustbenchmarkingquantumoptimization}. Due to the fundamentally different nature of quantum algorithms and the existing limitations in their implementation, accurately measuring their performance for comparison with classical algorithms is difficult.

Time to Solution (TTS) was used to make the comparison, which refers to the duration or computational effort required to find an optimal solution to a problem using a specific method. For classical exact algorithms such as brute force or branch-and-price, TTS is typically measured by the wall-clock time the algorithm is running (TTS=$t_{solve}$). However, when considering heuristic quantum algorithms, a more sophisticated method is needed to measure the TTS, given by Eq. \ref{eq:tts}.

\begin{equation}
    TTS(X)=\frac{t_{solve}}{M} \left( \frac{log(1-0.99)}{log(1-p^*)} \right)
    \label{eq:tts}
\end{equation}

In this equation, $p^*$ is the success probability and $T=t_{solve}/M$ represents the time needed to execute a single shot. 

\begin{figure}[h]
    \centering
    \includegraphics[width=0.65\linewidth]{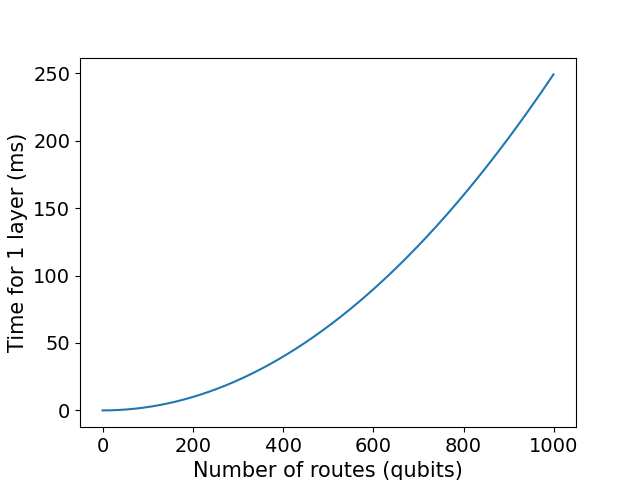}
    \caption[Time per layer as a function of the size of the problem]{Time per layer in the QAOA circuit as a function of the size of the problem.}
    \label{fig:time}
\end{figure}

When solving with QA, $T$ is given by the annealing time. For QAOA, it is appropriate to determine this time $T$ through the number of gates in the quantum circuit. In the worst-case scenario, the two-qubit gates are assumed to be implemented sequentially. Then, if one-qubit gates are estimated to need 50ns and two-qubit gates 500ns, the resulting time for the implementation of one layer takes the values represented in Figure \ref{fig:time}. This time has to be multiplied by the number of layers used in the quantum circuit.

\begin{figure}[h]
    \centering
    \includegraphics[width=0.8\linewidth]{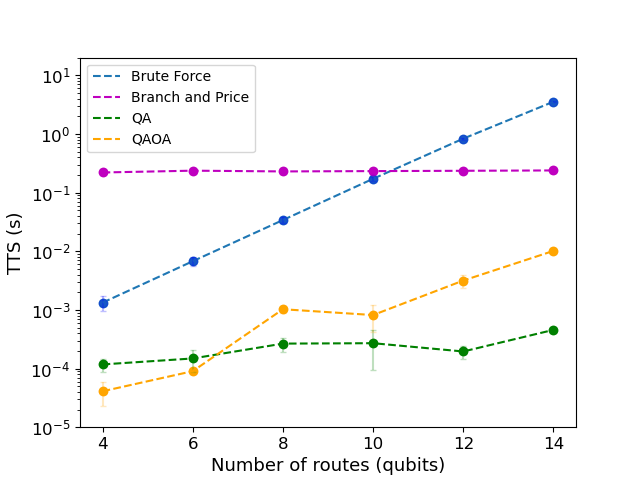}
    \caption[Scalability of the different approaches]{Scalability of the different approaches for instances ranging from 4 to 14 routes.}
    \label{fig:sc}
\end{figure}

Figure \ref{fig:sc} shows the TTS for different instances using different methods, which include brute force, branch-and-price \cite{barnhart1998branch} and QA \cite{yarkoni2022quantum}. It is limited to 14 routes because of the range where QAOA can operate.

Brute force exhibits a steep increase in TTS as the number of routes grows, making it impractical for larger instances due to its exponential complexity. In contrast, branch-and-price demonstrates exceptional scalability and efficiency, maintaining stable TTS across the range of problem sizes. 

Talking about the quantum approaches, QA, although exhibiting a slight increase in TTS with growing problem size, remains remarkably stable, demonstrating its effectiveness up to this point. On the other hand, the performance of QAOA appears less favorable in this context, with a significant increase in TTS as the problem size grows. It shows a slope slightly lower than that of brute force, but scales worse than both branch-and-price and QA. 

\begin{figure}[h!]
    \centering
    \includegraphics[width=1\linewidth]{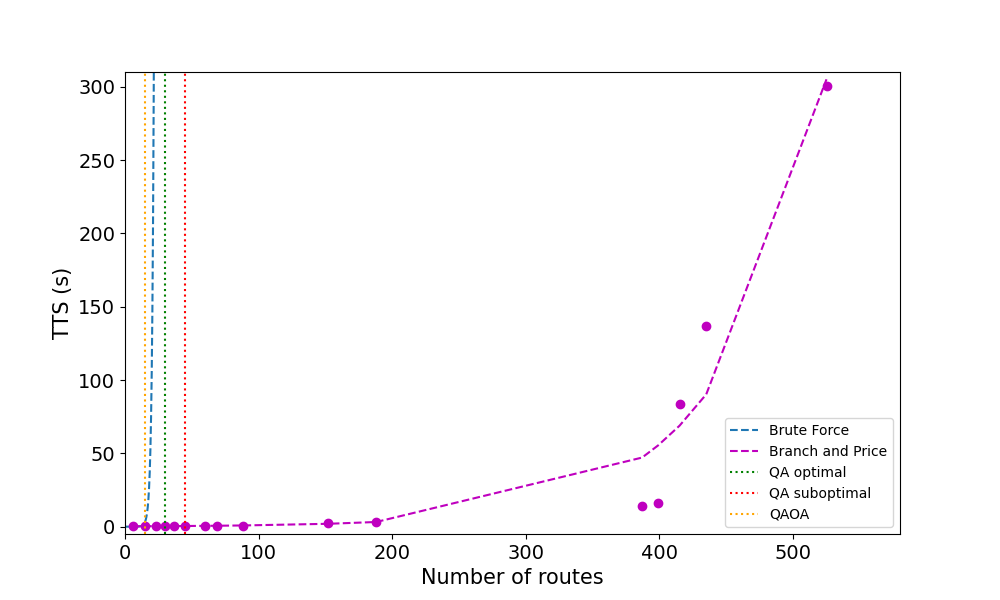}
    \caption{Current capacity of each approach.}
    \label{fig:curr}
\end{figure}

The challenge lies in the fact that within the range where quantum approaches can operate, classical algorithms like branch-and-price perform exceptionally well, providing solutions in very short times and demonstrating great scalability. However, as the size of the instances increases, there comes a point beyond which branch-and-price needs much more time, thus becoming inefficient. The true advantage of quantum approaches may become evident for these large instances, where classical algorithms struggle, but for the moment they are only able to address smaller instances. Figure \ref{fig:curr} illustrates that while branch-and-price can effectively manage a large number of routes before becoming inefficient, as indicated by the dotted line representing an exponential regression, quantum approaches are limited to a significantly smaller range, marked by the dotted vertical lines on the left.

\section{\label{sec:conc}Conclusions and Future Work}

In this work, QAOA has been applied to real-world instances of the TAP, a field that had not been extensively studied. After implementing the algorithm, improvements related to parameter initialization have been introduced through the INTERP strategy. Then, it has been observed that, when a sufficiently large number of layers is applied in the quantum circuit, the algorithm is capable of distinguishing the optimal from the suboptimal solutions for a given instance. Additionally, it has been found that the complexity of the problem increases as the connectivity in the route graph becomes higher.

Comparing classical and quantum approaches is challenging due to the different resources available, highlighting the need for advancements in quantum hardware to enable a more meaningful evaluation. 

Several enhancements can be introduced in future work to try to improve the scalability of the current QAOA implementation. One promising direction involves exploring alternative circuit ansatz, experimenting with different mixer Hamiltonians. Improving the classical optimization methods, for example using adaptive optimization strategies, is another strategy that could overcome current performance limitations. These and other potential improvements are discussed in detail in the work by Blekos et al., providing valuable insights for further refinement of the algorithm \cite{blekos2024review}.

Furthermore, testing the algorithm on real quantum hardware would be crucial for understanding the impact of noise and hardware constraints on performance. While the current work has focused on simulations, running QAOA on actual quantum devices could reveal important practical considerations. 

\begin{acknowledgments}

This paper has been supported by the  Government of the Basque Country through the research grant ELKARTEK KUBIT - Grant number KK-2024/00105 (Kuantikaren Berrikuntzarako Ikasketa Teknologikoa).

\end{acknowledgments}

\nocite{*}

\bibliography{references}% Produces the bibliography via BibTeX.

\end{document}